\documentclass[aps,prd,twocolumn,nofootinbib,notitlepage]{revtex4-1}
\usepackage{amsmath,amssymb,amstext,graphicx,color,xfrac,braket,accents,tabulary,booktabs,xcolor,bbold,hyperref,natbib}
\begin{document}

\thispagestyle{empty}

\newcommand{\be}{\begin{equation}}
\newcommand{\ee}{\end{equation}}
\newcommand{\bea}{\begin{eqnarray}}
\newcommand{\eea}{\end{eqnarray}}
\newcommand{\hst}{\widetilde{\mathcal{H}}} 
\newcommand{\iso}{\dot{=}}
\newcommand{\Dim}{\textrm{dim\,}}
\newcommand{\Tr}{\textrm{Tr\,}}
\newcommand{\hs}{\mathcal{H}} 
\newcommand{\E}{\widetilde{E}}
\newcommand{\thetat}{\tilde{\theta}}
\newcommand{\ham}{\widehat{H}}
\newcommand{\intham}{\widehat{H}_{\rm{int}}}
\newcommand{\selfham}{\widehat{H}_{\rm{self}}}
\newcommand{\dt}{\delta_t}
\newcommand{\trace}{\mathrm{Tr}}
\def\bra#1{\langle #1\rvert}
\def\ket#1{\lvert #1\rangle}
\newcommand{\draftnote}[1]{\textbf{\color{red}[#1]}}

\baselineskip=14pt

\title{Completely Discretized, Finite Quantum Mechanics}
\author{Sean M. Carroll}
\email{seancarroll@gmail.com}
\affiliation{Departments of Physics \& Astronomy and Philosophy, Johns Hopkins University,\\
Baltimore MD 21218 and Santa Fe Institute, Santa Fe, NM 87501}

\begin{abstract}
I propose a version of quantum mechanics featuring a discrete and finite number of states that is plausibly a model of the real world.
The model is based on standard unitary quantum theory of a closed system with a finite-dimensional Hilbert space.
Given certain simple conditions on the spectrum of the Hamiltonian, Schr\"odinger evolution is periodic, and it is straightforward to replace continuous time with a discrete version, with the result that the system only visits a discrete and finite set of state vectors.
The biggest challenges to the viability of such a model come from cosmological considerations.
The theory may have implications for questions of mathematical realism and finitism.
\end{abstract}

\maketitle

\section{Introduction}

Essentially all successful theories of modern physics -- classical mechanics, quantum mechanics, general relativity, quantum field theory -- are based on continuous mathematical structures such as smooth manifolds and vector spaces.
But it is well known that the continuum poses puzzles, both mathematically and philosophically \cite{sep-infinity}.
Physicists, accordingly, have long speculated about a discrete (countable, and possibly literally finite) structure that could underlie the physics of the real world.
The world looks pretty smooth, even to the precision of our best experimental probes, so the finite numbers involved would presumably have to be very large by ordinary standards, enough to make such dynamics well-approximated by continuum theories.
But at a foundational level, there is a significant conceptual difference between ``infinite" and ``really big."
It is therefore worth taking the feasibility of a simple finite model of physics seriously.

One route toward this ambition is to simply start from scratch with some kind of discrete structure, and ask whether it might conceivably map onto the real world.
This strategy has been pursued, for example, in the recent Wolfram Physics Project \cite{wolfram2020}.
But the hoped-for connection to known physics remains tenuous at this point.

Another route is to stay within the framework of quantum mechanics, which currently is the most fundamental theory we have.\footnote{Discussions of fundamental theories often revolve around approaches to unification and quantum gravity, such as superstring theory. But for the most part all of these approaches live within the standard framework of quantum mechanics; they are specific models under that broad umbrella, rather than alternatives to it.}
Quantum mechanics has a reputation for involving, well, quanta.
Energies and other observables can (under the right circumstances) take on discrete allowed values, with stochastic ``jumps'' between them.
Furthermore, we can combine the fundamental units of the speed of light $c$, Newton's gravitational constant $G$, and Planck's quantum constant $\hbar$ to form a fundamental unit of length, the Planck length,
\be
  L_p = \sqrt{\frac{\hbar G}{c^3}} = 1.6\times 10^{-32}~\textrm{cm,}
\ee
and a corresponding Planck time,
\be
  t_p = \sqrt{\frac{\hbar G}{c^5}} = 5.4\times 10^{-44}~\textrm{sec.}
\ee
From this observation, it is natural to wonder whether spacetime itself is fundamentally discrete on this scale, with some finite number of degrees of freedom per Planck volume $L_p^3$. (Henceforth we set $\hbar = c = 1$.)

However, neither of these notions -- the existence of a finite Planck scale, or the apparent discreteness of quantum jumps -- represents a fully discrete version of physics.
For one thing, the idea that spacetime is discrete is perhaps vaguely suggested, but certainly not directly implied, by anything we know for sure about quantum mechanics, gravity, or Planckian numerology.
We can certainly imagine models in which spacetime is continuous and yet the Planck scale plays some kind of role, as in the most straightforward understanding of string theory or the AdS/CFT correspondence \cite{Maldacena:1997re}. 
Models in which spacetime is straightforwardly discrete are under active investigation \cite{Surya_2019}, but it remains unclear whether this is the right way forward.

More importantly, quantum mechanics itself is not discrete, even if spacetime is.
In particular, the discreteness associated with quantum jumps is apparent, not fundamental.
The theory itself is built on the concept of a quantum state' $\ket{\psi}$, which is an element of a Hilbert space.
Hilbert spaces are continuous vector spaces, in which linear combinations of elements (with complex coefficients) are also elements.
So one needs the continuum to describe them.
In many theories -- including any quantum field theory, or indeed in the quantum theory of an ordinary non-relativistic particle -- the relevant Hilbert space is infinite-dimensional.
For that matter, the fundamental evolution equation for the state is the Schr\"odinger equation,
\be
\hat{H}\left|\psi\right\rangle=i\frac{\partial}{\partial t}\left|\psi\right\rangle,
\label{schr}
\ee
where $\hat{H}$ is the Hamiltonian operator, whose specific form defines the dynamics of the theory.
This is a continuous differential equation, not a discrete difference equation.
The time parameter takes on any real value.

For all of these reasons, conventional quantum theories are not candidates for the kind of simple, discrete theory we wish to consider.
We need to somehow modify the quantum formalism.
In this paper I point out that the necessary modification is extremely minor.
In the right kind of quantum theories (i.e. with an appropriate Hamiltonian), we can straightforwardly discretize the time evolution in such a way that only a discrete, finite set of states ever occur.
In that case the continuum nature of Hilbert space is irrelevant, since physical states only occupy a discrete lattice within that space.
(For other investigations of discrete Hilbert space, in slightly different forms, see \cite{Buniy2005,Palmer2022}.)
I then ask whether such a theory could plausibly match the real world.
I argue for a tentative affirmative, although there are nontrivial challenges associated with cosmological evolution.

\section{Discretized Quantum Mechanics}

Let us think in general terms about a quantum system governed by the Schr\"odinger equation (\ref{schr}).
In textbook treatments of quantum theory, one often starts with a classical system governed by some Hamiltonian $H(x,p)$ depending on some generalized positions $x$ and momenta $p$, and then ``quantizes" it. 
One way of doing this is to construct a complex-valued wave function of configuration space $\psi(x)$, then representing momentum as a differential operator $\hat{p} = -i \partial/\partial x$.
Plugging this into the Hamiltonian yields a Hamiltonian operator $\hat{H}$ that appears in the Schr\"odinger equation.
In this case, the abstract state vector $\ket{\psi}$ is represented by the function $\psi(x)$.

However, this process is somewhat misleading.
The quantization procedure, considered as a map from classical to quantum theories, is neither one-to-one nor onto, nor even a well-defined function.
One classical theory can correspond to multiple quantum theories, two different classical theories can correspond to the same quantum theory, and some quantum theories have no classical precursors.
Furthermore, it is not necessary.
It is useful for human beings to start with a classical model and quantize it, but nature presumably doesn't work that way; it is quantum from the start, and classicality emerges in an appropriate limit.

For our purposes, then, a ``quantum theory'' is specified by a Hilbert space $\mathcal H$ (that is, a complex normed vector space of fixed dimensionality $D$) and an Hermitian Hamiltonian operator.
When the dimensionality of $\mathcal H$ is uncountably infinite, it is also necessary to specify a set of observables; but we are going to be interested in finite $D$, so that any Hermitian operator is an observable.
Considerations of quantum gravity suggest that our observable universe is described by a finite-dimensional Hilbert space \cite{Bao:2017rnv}, with $D\sim 10^{10^{122}}$.
Whether or not the entire physical universe is described by a finite-dimensional quantum theory is an open question; for our present purposes we assume it can be.

We need to choose a way to specify the Hamiltonian.
For that purpose we note that we can define energy eigenstates  $\{|k\rangle\}$ satisfying 
\be
  \widehat{H}\left|k\right\rangle=E_k\left|k\right\rangle,
  \label{eigenbasis}
\ee
where each eigenvalue $E_k$ is interpreted as the energy of state $\ket{k}$.
For convenience we will assume that the $E_k$ are all different numbers (a non-degenerate Hamiltonian).
Then the states $\ket{k}$ form an orthonormal basis for the vector space $\mathcal H$.
It is in this basis that the Hamiltonian is simply a diagonal matrix whose entries are the real numbers $\{E_k\}$.
Listing these eigenvalues -- the ``spectrum'' of the Hamiltonian -- completely specifies our quantum theory.

One nice thing about the energy eigenbasis (aside from being the only uniquely-defined basis that depends only on the Hamiltonian) is that it is straightforward to write down an exact solution to the Schr\"odinger equation.
Combining (\ref{schr}) with (\ref{eigenbasis}) shows that a system that is originally in an energy eigenstate, $\ket{\psi(0)} = \ket{k}$, will evolve as
\be
  \ket{\psi(t)} = e^{-iE_kt}\ket{k}.
  \label{energyevol}
\ee
Physically, this can be interpreted to say that the system simply remains in the original energy eigenstate.
That's because an overall phase factor has no physical reality, since according to the Born Rule the probability of an observational outcome is equal to the norm-squared of the associated component of the wave function.
That is, the probability of observing a system in state $\ket{\psi}$ to be in a state $\ket{a}$ is given by constructing the dual ket $\langle a|$ and calculating
\be
  P(a) = |\langle a | \psi\rangle|^2.
\ee
This number is unaffected by a phase rotation $\ket\psi \rightarrow e^{i\theta}\ket\psi$, so that an overall phase factor is irrelevant.

But the \emph{relative} phases between different components do matter.
Given any initial state $\ket{\psi(0)}$, we can always write it as a superposition of basis elements, in this case energy eigenstates:
\be
  \ket{\psi(0)} = \sum_{k}{\alpha_k\left|k\right\rangle},
\ee
where the $\{\alpha_k\}$ are real-valued parameters, with $k\in \{0, 1, 2, \ldots D-1\}$.
(We have absorbed any initial phases into the definition of the basis states $\ket k$.)
Then, from the linearity of the Schr\"odinger equation, the time-evolution of this superposition is just the superposition of time-evolved states of the form (\ref{energyevol}).
This can be written as
\be
\left|\psi\left(t\right)\right\rangle=\sum_{k=0}^{D-1}{\alpha_ke^{-i\theta_k(t)}\left|k\right\rangle},
\label{psi-evol}                         
\ee
where
\be
  \theta_k(t) = E_kt.
\ee
The amplitudes $\{\alpha_k\}$ remain constant.

One subtlety is that we can still perform an arbitrary shift on all the energy eigenvalues, $\{E_k\} \rightarrow \{ E_k + c\}$, since that introduces an irrelevant overall phase.
(Even a time-dependent overall phase is still physically unobservable.)
Let us assume that we are working with a finite-dimensional Hilbert space, and without loss of generality let us order the eigenvalues from lowest to highest, so that $E_k < E_{k+1}$.
Then we can define shifted energies by extracting the lowest eigenvalue,
\be
  \E_k = E_k - E_0,
\ee
and write the state as
\be
\left|\psi\left(t\right)\right\rangle= \alpha_0 |0\rangle + \sum_{k=1}^{D-1}{\alpha_k e^{-i\thetat_k(t)}\left|k\right\rangle}, 
  \label{finalstate}                  
\ee
where
\be
  \thetat_k(t) = \E_k t.
\ee
The physically irrelevant overall phase $e^{-iE_0t}$ has been ignored.

Staring at this, we see that Schr\"odinger evolution is quite simple. 
All of the time-dependence appears in a set of phases that chug forward linearly in time:
\be
\thetat_k\left(t+\Delta t\right)=\thetat_k\left(t\right)+\E_k\Delta t.                                                   
\ee
The phases themselves are periodic variables, since $\thetat_k\equiv\thetat_k+2\pi$. 
The set of all phases is therefore a product of $D-1$ circles, i.e. a $(D-1)$-dimensional torus, and the entirety of Schr\"odinger evolution can be thought of as the set of phases moving in a straight line on this torus.

If the ratio of any two (shifted) energies is a rational number, $\forall (k,j): \E_k/\E_j \in \mathbb{Q}$, we say that the set of energies is commensurable.
If the energies are incommensurable, then the evolution of the phases will never come precisely back to its starting point; instead it will fill out some (possibly lower-dimensional) torus.
But if the energies are all commensurable, then the evolution will be exactly periodic, describing a one-dimensional closed curve (topologically a circle) within the torus of phases.
In that case we can uniquely write
\be
  \E_k = p_k \epsilon,
  \label{pk}
\ee
where $\epsilon$ is some fixed unit of energy and $\{p_k\}$ is a set of coprime integers.
(A set of integers is coprime if there is no common divisor of the set as a whole.)
Then after a time
\be
  T_\mathrm{recur} = \frac{2\pi}{\epsilon},
\ee
each phase will increment $\thetat_k \rightarrow \thetat_k + 2\pi p_k$, and the state vector will return to where it started.

\begin{figure}[h]
\begin{center}
\includegraphics[width=0.45\textwidth]{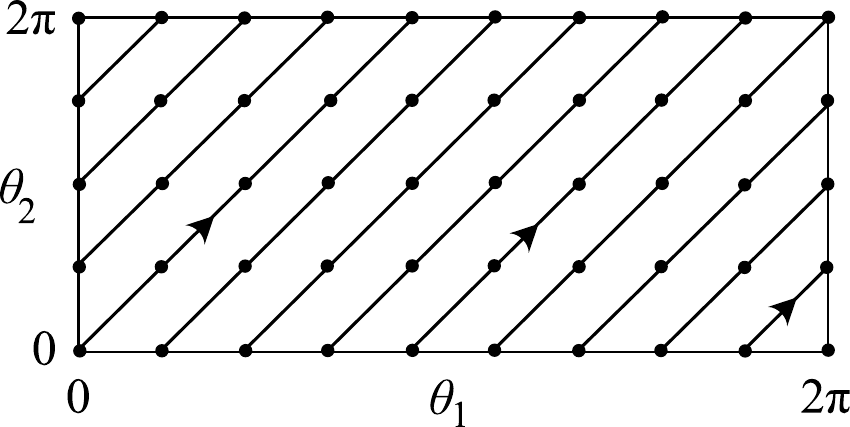}
\caption{Evolution in a two-dimensional torus of phases. For this example we have $p_1=4, p_2=9, N=36$.
The right-hand side of the drawn rectangle is identified with the left-hand side, and the top is identified with the bottom, forming a torus. Ordinary smooth Schr\"odinger evolution is portrayed by the diagonal line, and discrete timesteps are shown as solid dots.}
\end{center}
\end{figure}

Let us assume that our system is indeed described by a set of commensurable energies.
While conventional quantum theory describes smooth evolution, in this case such evolution is readily made discrete.
Let $N$ be the least common multiple of the integers $\{p_k\}$,
\be
  N = \mathrm{LCM}(\{p_k\}).
\ee
Then define a fundamental timestep
\be
  \dt = \frac{T_\mathrm{recur}}{N} = \frac{2\pi}{N\epsilon}.
\ee
Note that $N$ is likely to be extremely large, being the least common multiple of $D-1$ integers, where the dimension of Hilbert space is plausibly $D \sim 10^{10^{122}}$.
So the timestep $\dt$ is presumably much shorter than the Planck time.
When considering the coefficient of $\ket k$, we note that after $N/p_k$ timesteps, the phase increments by
\bea
  \thetat_k &\rightarrow& \thetat_k + \E_k \left(\frac{N}{p_k}\right)\dt,\\
  &=& \thetat_k + (p_k\epsilon) \left(\frac{N}{p_k}\right)\left(\frac{2\pi}{N\epsilon}\right)\\
  &=& \thetat_k + 2\pi.
\eea
After $N$ total timesteps $\dt$, all of the phases will increment by multiples of $2\pi$, so the state returns to its original form.

The overall evolution of this discretized model is periodic in time, with precisely $N$ distinct states realized along the way.
The evolution of the universe is cyclic, repeating after $N$ timesteps.
Note that this is not ordinary quantum mechanics, where time is continuous; it is a modification thereof.
But if $D$ and $N$ are sufficiently large, it might reasonably be indistinguishable from ordinary quantum theory.
The evolution is also perfectly reversible, as in conventional Schr\"odinger evolution, as distinct from many cellular automata or other attempts at discretizing physics.

We should also ask about the amplitudes $\alpha_k$ in (\ref{finalstate}); aren't they arbitrary real numbers?
In ordinary quantum mechanics they are, up to a normalization
\be
  \sum_k (\alpha_k)^2 = 1.
  \label{normalization}
\ee
That would imply that for each $k$, $|\alpha_k| \leq 1$.
But this is somewhat misleading.
The numerical value of the normalization in (\ref{normalization}) is not physically meaningful.
We simply choose it to be 1, since then then the probability formula takes on a particularly simple form.
But all that really matters is that the norm of the wave function is constant; i.e. that the right-hand side of (\ref{normalization}) is some fixed number.
As long as the set of $\alpha_k$'s are commensurable, we can convert them all to integers (perhaps very large ones) by multiplying by their least common denominator $L$, defining $\tilde\alpha_k = L\alpha_k$, and writing the state vector with amplitudes $\tilde\alpha_k$.
Then we will have $\sum_k |\tilde\alpha_k|^2 = L^2$, which is unconventional but completely legitimate.
So once again we only really need a finite set of integers.

\section{Connecting to Reality}

Such a simple model -- eternal linear evolution on a discrete torus -- might seem too trivial to capture something like the real world, which we typically describe in terms of complicated things like quantum fields and curved spacetime. 
But it is plausible that those familiar pieces of ontology are emergent, appearing as parts of a higher-level description from a single quantum state vector evolving unitarily (\cite{Carroll:2018rhc,Carroll2021} and references therein). 
While the Schr\"odinger equation itself is linear, the nonlinearities familiar from our everyday experience can arise as part of an emergent classical limit.

We have been implicitly relying on the Everettian approach to quantum mechanics \cite{wallace2012emergent}.
In this approach, the quantum state is taken as a comprehensive representation of reality (unlike hidden-variable theories), and it always evolves according to the Schr\"odinger equation (unlike objective-collapse models).
The apparent collapse of the wave function during quantum measurement is explained by the process of decoherence, which splits the quantum state into mutually non-interacting branches that subsequently evolve as independent worlds.
In appropriate circumstances, these worlds can be approximated well by classical evolution.
(Exceptions occur when stochastic quantum measurement outcomes are amplified to macroscopic differences, such as when a person makes a decision on the basis of the outcome of a quantum measurement.)

It is possible, however, to be an Everettian and invoke more than just a vector in Hilbert space as part of one's fundamental ontology. 
As mentioned, physicists traditionally construct quantum theories by quantizing classical ones, so that wave functions are thought of as square-integrable complex functions of the configuration space (or momentum space, or equivalent) of the classical precursor.
Such a precursor might be a set of fields defined on spacetime, or a set of particle positions.
Then one might contemplate the existence of a ``preferred'' classical precursor, which remains part of the ontology even after the transition to quantum mechanics.
The proposal dubbed ``wave function realism," for example, suggests that the quantum state be thought of specifically as a function of the configuration space of particles in three-dimensional physical space \cite{Albert1996-ALBEQM}.

The philosophy here is different. 
We are taking the quantum state itself, a vector evolving in Hilbert space, as the sole element of our ontology. 
To connect to known features of the world, therefore, requires a kind of reverse-engineering: going from an extremely minimal set of data (energy eigenvalues, amplitudes of the actual wave function of the universe) to the rich world of our experience.
The hope is that this minimal data corresponds uniquely to an emergent quasiclassical structure.
Determining whether this might work is an ongoing project.

We won't review the entirety of this program here, but we can mention the main ideas.
One step is to pinpoint a classical limit.
This involves finding a preferred factorization of Hilbert space, in which it can be written as a tensor product of ``system'' and ``environment'' factors, so that the system behaves classically as it is monitored by the environment.
This can be done by looking for factorizations that permit the system to remain localized with respect to certain observables (which are then interpreted as ``positions in space'') and relatively unentangled with the environment \cite{Carroll:2020gme}.
Of course the system of our observed world takes a specific form: a four-dimensional spacetime inhabited by local quantum fields and curvature that responds to energy-momentum.
It was recently established that the spectrum of the Hamiltonian is enough to pick out the appropriate notions of locality and dimensionality, when such exist \cite{Cotler:2017abq}.
By assuming that distances within such an emergent structure are inversely related to the entanglement of vacuum degrees of freedom, an emergent geometry can be defined that plausibly obeys Einstein's equation of general relativity \cite{Cao:2016mst,Cao:2017hrv}.
It remains to extract the known fermions, gauge fields, and Higgs sector that we know from the Standard Model of particle physics, but proposals such as ``string net condensation'' may prove to be useful in this regard \cite{Levin:2004mi}.

Much remains to be established, but it seems plausible that physics resembling that of our familiar world can be reverse-engineered, at least in principle, from a minimal recipe of a quantum state and eigenvalues of a Hamiltonian.
At least, in conventional quantum theory; the special finite discrete theory considered here raises unique questions, especially as concerns cosmology.

A finite-dimensional Hilbert space is compatible with an emergent spacetime that takes the form of de~Sitter space, a cosmological spacetime with a positive vacuum energy.
Classically, de~Sitter can be foliated into spacelike surfaces with either finite or infinite volume.
But it is possible that this classical picture is overly naive. 
In the quantum context, what matters is that de~Sitter space has a horizon, with a Gibbons-Hawking entropy proportional to its area measured in Planck units.
De~Sitter is also a maximum-entropy state \cite{Carroll:2017kjo}, so that entropy should be approximately equal to the logarithm of the dimensionality of Hilbert space inside the horizon \cite{Banks:2000fe}.
This is the origin of the idea that the dimensionality of the Hilbert space describing our observable universe is
\be
  D \approx 10^{10^{122}}.
\ee
And in the context of horizon complementarity, it is plausible that the degrees of freedom inside and on the horizon are all the degrees of freedom there are, so that only a finite-dimensional Hilbert space is required for the theory as a whole \cite{Nomura:2011dt,Bousso:2011up}.
We can furthermore sketch a picture in which emergent space expands from a Big Bang beginning to a larger cosmological spacetime, by positing that a collection of initially unentangled degrees of freedom are gradually brought into an entangled group defining space and its ingredients at a fixed time \cite{circuitcosmo}.

Cosmology becomes potentially problematic for models of discretized finite-dimensional Hilbert space because the overall time-evolution is necessarily periodic.
We therefore have recurrence: anything that happens once will happen an infinite number of times in the future, and has happened an infinite number of times in the past.
By itself this is okay, but there is also the prospect of Boltzmann-Brain-like occurrences: appearances of observers that randomly fluctuate into existence, rather than arising as the result of ordinary thermodynamically sensible evolution from an initial low-entropy state \cite{Dyson2002,Albrecht_2004,Carroll:2008yd}.

The problem with such fluctuations is not the existence of literal Boltzmann Brains -- minimal fluctuations that satisfy some particular conditions for making an ``observer," presumably representing primitive neural configurations surrounded by empty space.
The problem is that even when we conditionalize on observers that are macroscopically indistinguishable from ourselves, including all of our impressions of our environments, it is still overwhelmingly likely for them to be random fluctuations rather than ordinary observers.
As a result, everything in the brains of such observers would represent a random fluctuation, including all supposed knowledge of the rules of reasoning and data about the empirical world.
It would be unjustified for such observers to conclude anything reliable about reality -- the situation is cognitively unstable \cite{Carroll2017}.

Does the kind of periodic evolution considered here lead to a Boltzmann-fluctuation problem?
One's first guess would be ``yes," but the situation is not completely clear.
We can imagine two ways out, though neither of them is truly compelling at the moment.

The first way out is to have the periodic evolution be non-ergodic in a noticeable way, avoiding the appearance of Boltzmann fluctuations like ourselves.
In a finite classical thermal system that lasts infinitely long, given some coarse-graining into macrostates with entropy given by Boltzmann's formula $S=k\log W$ (with $W$ the volume of the macrostate), we expect the system to spend most of its time in high-entropy equilibrium macrostates.
There will be occasional fluctuations downward in entropy by an amount $\Delta S$, with probability $P \propto e^{-\Delta S}$.
If the evolution is ergodic, at least within some representative subspace, most achieved states satisfying any particular local condition (such as the existence of a specified kind of observer) will otherwise be as high-entropy as possible. 

The generic expectation would be that recurrent quantum evolution would exhibit analogous behavior, with the system spending most of its time in or near an equilibrium macrostate, with occasional fluctuations downward in entropy.
That is an unacceptable model of the real world, due to the cognitive-instability problem.
Even if it were true, we would never have legitimate grounds for believing it.
But we can imagine a non-generic, fine-tuned kind of evolution, featuring a single enormous downward fluctuation in entropy with the rest of the evolution staying in a high-entropy state with no fluctuations.
(Contrary to informal talk about ``quantum fluctuations," a truly stationary quantum state does not feature any dynamical processes creating lower-entropy configurations \cite{Boddy_2016}. See \cite{Lloyd:2016ahu} for an alternate perspective.)
The large entropy decrement would be identified with the Big Bang in our observed universe.
The classical interpretation of the resulting evolution would be expansion from a Big Bang singularity, dilution and cooling to an ultra-long-lived state resembling empty de~Sitter space, followed by a Big Crunch with decreasing entropy, after which the sequence repeats forever.
This scenario is entirely speculative, but would allow for periodic evolution without Boltzmann fluctuations, and is therefore worth further investigation. 

The second way out is if the time evolution is not truly periodic because time itself is emergent.
This possibility has been discussed in a variety of contexts \cite{Page:1983uc,Banks:1984cw,Albrecht:2007mm,Rovelli:2009ee,Giovannetti:2015qha,Marletto:2016gwv,Singh:2020kdu}.
There are different ways to implement the program, but the general idea is to factor Hilbert space into a factor representing a ``clock'' and another representing the rest of the universe.
With an appropriate entanglement structure, the state of the rest of the universe can be entangled with clock states of definite ``time'' so that the universe appears to evolve under an emergent Hamiltonian with respect to the clock readings.

If time is emergent and Hilbert space is finite-dimensional, there will only be a finite number of ticks on the effective clock.
That provides an obvious way to avoid eternal recurrence and Boltzmann fluctuations: allow for sufficient ticks to describe evolution from a low-entropy beginning to ultimate thermal equilibrium, but not enough time to render substantial fluctuations inevitable.
It is unclear at this point why the state of the universe should be arranged in this way, but it appears to be a viable possibility.

If neither of these options can be made more concrete and successful, the idea of an eternal universe with a finite-dimensional Hilbert space would have to be discarded.
If one can be made to work, then we can imagine a truly discrete and finite formulation of realistic physics that is not too much of a departure from conventional quantum mechanics.

\section{Discussion}

The model presented here, notwithstanding the cosmological issues just discussed, potentially represents a phenomenologically realistic physics that is completely discrete and finite, rather than relying on continuous quantities in the definition of time evolution or the space of states.
We can ask whether this actually helps avoid the physical and philosophical puzzles associated with infinity, with somewhat ambiguous answers.

On the physical side, continuum-based theories can run into problems with singularities, where purportedly physical quantities become infinite -- curvature singularities in general relativity, or divergent scattering amplitudes in quantum field theory.
If such theories are understood as smooth approximations to an underlying discrete structure such as that presented here, those problems instantly evaporate.
Behavior is well-defined and non-singular for the entirety of its (periodic) evolution; there is nothing to become infinite.

The philosophical puzzles are more subtle.
Having a finite number of states helps avoid some of the more obvious conceptual problems associated with infinity \cite{sep-infinity}.
But the foundational mathematical tools we use to describe physical systems, including Peano Arithmetic (PA) and Zermelo-Fraenkel (ZF)  set theory, involve infinite quantities whether or not they correspond directly to physical structures.
This is related (but not identical) to the fact that PA and ZF are powerful enough to be subject to G\"odel's incompleteness theorems, and therefore be unable to prove their own consistency or completeness.
This has led to the suggestion that the need to trust in the consistency of our physical descriptions leads us to a form of mathematical realism, in which we assert such consistency even though it is unprovable \cite{putnam1975mathematical}.
One can contrast this view with one in which only the physical world is ``real," and mathematics is simply a way of talking about that world \cite{CarrollManuscript-CARRRK}, without a distinct reality of its own.
This position is defensible even if we need PA to describe the world, but it becomes much more intuitive if we are able to completely capture physical reality using mathematics that is complete and decidable, such as Presburger arithmetic (PrA) \cite{presburger1929uber,haase}.

Is Presburger arithmetic all we need to capture completely discretized quantum mechanics?
PrA describes the operations of addition and order on the natural numbers, but not multiplication or exponentiation.
Modular addition can be included, therefore PrA suffices to capture the evolution on a toroidal lattice that is used in this model.
The evolution of energy eigenstates is traditionally expressed as $e^{-iE_k}|k\rangle$, which involves the exponential of an imaginary number, or equivalently a sum of trigonometric functions.
But that is presumably a matter of convenience; we could easily describe that evolution directly on the lattice, without relying on exponentials or trig functions.

We are still left, however, with the fact that the quantum state as a whole includes the amplitudes $\tilde\alpha_k$, which appear as coefficients of the basis vectors in the expansion of the state vector.
Even if the $\tilde\alpha_k$s themselves are integers, it seems necessary to multiply them by the basis vectors, which is an operation that PrA is unable to describe.
We tentatively conclude that discretized quantum mechanics still requires a mathematical formalism more powerful than PrA, but it remains possible that it is simply a matter of transforming variables appropriately.

One could go further and ask whether a model like the one examined here could be compatible with finitism in the philosophy of mathematics \cite{ye2011strict}.
Finitists reject the use of infinite sets and unbounded quantification.
A common light-hearted objection asks whether this implies the existence of a largest possible number, and what happens when you reach it?
In a finite physical model, there is no largest possible number, but there is a largest possible number of states that the universe can be described by. 
While there is no logically necessary connection between finitism as an approach in the philosophy of mathematics and the actual physical structure of the world, the possibility that the world is itself strictly finite might make the approach seem more palatable.

It is interesting that sone of the most basic questions we can ask about physical reality, whether it is discrete or continuous and whether it is finite or infinite, remain unanswered.
The model considered here offers one way of making such questions more concrete.

\section*{Acknowledgements}
It is a pleasure to thank Justin Clarke-Doane for helpful comments.

\bibliographystyle{utphys}
\bibliography{discreteQM}
\end{document}